\documentclass[twoside,reqno]{mbook}
\usepackage{epsfig,cite}
\usepackage{amssymb,amsmath}
\usepackage{times}
\begin{document}

\sloppy \raggedbottom

 \setcounter{page}{1}

\newpage
\setcounter{figure}{0}
\setcounter{equation}{0}
\setcounter{footnote}{0}
\setcounter{table}{0}
\setcounter{section}{0}



\title{The Nuclear Force Problem: Is the Never-Ending Story Coming to an End?}

\runningheads{The Nuclear Force Problem}{R.~Machleidt}

\begin{start}
\author{R.~Machleidt}{}

\address{Department of Physics, University of Idaho, Moscow, Idaho, U.S.A.}{}

\begin{Abstract}
The attempts to find the right (underlying) theory for the
nuclear force have a long and stimulating history.
Already in 1953, Hans Bethe stated that "more man-hours
have been given to this problem than to any other scientific
question in the history of mankind". In search for the
nature of the nuclear force, the idea of sub-nuclear
particles was created which, eventually, generated the 
field of particle physics. I will review this productive 
history of hope, error, and desperation. 
Finally, I will discuss recent ideas
which apply the concept 
of an effective field theory to low-energy QCD. There are 
indications that this concept may provide the right framework 
to properly understand nuclear forces.
\end{Abstract}
\end{start}

\section{Historical perspective}

The theory of nuclear forces has a long history (cf.\ Table~1).
Based upon the seminal idea by Yukawa~\cite{Yuk35}, 
first field-theoretic attempts
to derive the nucleon-nucleon (NN) interaction
focused on pion-exchange.
While the one-pion exchange turned out to be very useful
in explaining NN scattering data and the properties
of the deuteron~\cite{Sup56}, 
multi-pion exchange was beset with
serious ambiguities ~\cite{TMO52,BW53}.
Thus, the ``pion theories'' of the 1950s
are generally judged as failures---for reasons
we understand today: pion dynamics is constrained by chiral
symmetry, a crucial point that was unknown in the 1950s.

\begin{table}[h]
\caption{Seven Decades of Struggle:
The Theory of Nuclear Forces} 
\begin{tabular*}{\textwidth}{@{\extracolsep{\fill}}cccc}
\hline
\hline
\\
   & 
  \bf 1935   &
\bf Yukawa: Meson Theory &
\\
\\
\hline
     &      &
{\it The ``Pion Theories''}
\\
 & \bf 1950's &
One-Pion Exchange: o.k.
\\
  &         &
Multi-Pion Exchange: disaster
\\
\hline
  &         & 
Many pions $\equiv$ multi-pion resonances:
\\
 & \bf 1960's & 
{\boldmath $\sigma$, $\rho$, $\omega$, ...}
\\
  &         & 
The One-Boson-Exchange Model
\\
\hline
  &         &
            Refine meson theory:
\\
 & \bf 1970's & 
Sophisticated {\boldmath $2\pi$} exchange models
\\
  &         & 
(Stony Brook, Paris, Bonn)
\\
\hline
  &         & Nuclear physicists discover
\\
 & \bf 1980's &  {\bf QCD}
\\
  &        & Quark Cluster Models
\\
\hline
  &        &
Nuclear physicists discover {\bf EFT}
\\
 & \bf 1990's &
Weinberg, van Kolck
\\
 & \bf and beyond &
{\bf Back to Meson Theory!}
\\
  &           &
{\it But, with Chiral Symmetry}
\\
\hline
\hline
\end{tabular*}
\end{table}

Historically, the experimental discovery of heavy 
mesons~\cite{Erw61} in the early 1960s
saved the situation. The one-boson-exchange (OBE)
model~\cite{OBEP,Mac89} emerged which is still the most economical
and quantitative
phenomenology for describing the 
nuclear force~\cite{Sto94,Mac01}.
The weak point of this model, however, is the scalar-isoscalar
``sigma'' or ``epsilon'' boson, for which the empirical
evidence remains controversial. Since this boson is associated
with the  correlated (or resonant) exchange of two pions,
a vast theoretical effort that occupied more than a decade 
was launched to derive the 2$\pi$-exchange contribution
to the nuclear force, which creates the intermediate 
range attraction.
For this, dispersion theory as well as 
field theory were invoked producing  the
Paris~\cite{Vin79,Lac80} and the Bonn~\cite{Mac89,MHE87}
potentials.

The nuclear force problem appeared to be solved; however,
with the discovery of quantum chromo-dynamics (QCD), 
all ``meson theories'' were
relegated to models and the attempts to derive
the nuclear force started all over again.

The problem with a derivation from QCD is that
this theory is non-perturbative in the low-energy regime
characteristic of nuclear physics, which makes direct solutions
impossible.
Therefore, during the first round of new attempts,
QCD-inspired quark models~\cite{MW88} became popular. 
These models are able to reproduce
qualitatively and, in some cases, semi-quantitatively
the gross features of the nuclear 
force~\cite{EFV00,Wu00}.
However, on a critical note, it has been pointed out
that these quark-based
approaches are nothing but
another set of models and, thus, do not represent any
fundamental progress. Equally well, one may then stay
with the simpler and much more quantitative meson models.

A major breakthrough occurred when 
the concept of an effective field theory (EFT) was introduced
and applied to low-energy QCD~\cite{Wei79}.

Note that the QCD Lagrangian for massless up and down quarks
is chirally symmetric, i.~e., it is invariant under
global flavor 
$SU(2)_L \times SU(2)_R$ equivalent to
$SU(2)_V \times SU(2)_A$ (vector and axial vector)
transformations. The axial symmetry is spontaneously broken
as evidenced in the absence of parity doublets in the
low-mass hadron spectrum. This implies the existence 
of three massless Goldstone bosons which are identified
with the three pions ($\pi^\pm, \pi^0$).
The non-zero, but small, pion mass is a consequence of
the fact that the up and down quark masses are not
exactly zero either (some small, but explicit symmetry breaking).
Thus, we arrive at a low-energy scenario 
that consists 
of pions and nucleons interacting via a force
governed by spontaneously broken approximate chiral
symmetry. 

To create an effective field theory describing this scenario,
one has to write down the most general Lagrangian consistent
with the assumed symmetry principles, particularly
the (broken) chiral symmetry of QCD~\cite{Wei79}.
At low energy, the effective degrees of freedom are pions and
nucleons rather than quarks and gluons; heavy mesons and
nucleon resonances are ``integrated out''.
So, the circle of history is closing and we are back to Yukawa's meson theory,
except that we have learned to add one important refinement to the theory:
broken chiral symmetry is a crucial constraint that generates
and controls the dynamics and establishes a clear connection
with the underlying theory, QCD.

It is the purpose of the remainder of this paper to describe
the EFT approach to nuclear forces in more detail.

\section{Chiral perturbation theory and the hierarchy
of nuclear forces}

The chiral effective Lagrangian 
is given by an infinite series
of terms with increasing number of derivatives and/or nucleon
fields, with the dependence of each term on the pion field
prescribed by the rules of broken chiral symmetry.
Applying this Lagrangian to NN scattering generates an unlimited
number of Feynman diagrams.
However, Weinberg showed~\cite{Wei90} that a systematic expansion
exists in terms of $(Q/\Lambda_\chi)^\nu$,
where $Q$ denotes a momentum or pion mass, 
$\Lambda_\chi \approx 1$ GeV is the chiral symmetry breaking
scale, and $\nu \geq 0$ (cf.\ Figure~1).
This has become known as chiral perturbation theory ($\chi$PT).
For a given order $\nu$, the number of terms is
finite and calculable; these terms are uniquely defined and
the prediction at each order is model-independent.
By going to higher orders, the amplitude can be calculated
to any desired accuracy.

\begin{figure}[h]
\vspace*{-0.5cm}
\hspace*{0.8cm}
\psfig{file=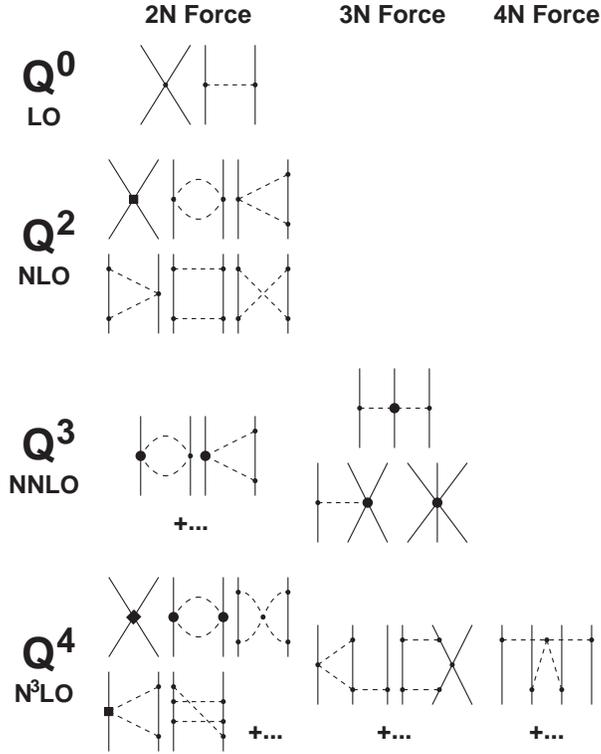,width=90mm}
\vspace*{-0.5cm}
\caption{Hierarchy of nuclear forces in $\chi$PT. 
Solid lines represent nucleons and dashed lines pions. 
Further explanations are given in the text.}
\end{figure}

Following the first initiative by Weinberg \cite{Wei90}, pioneering
work was performed by Ord\'o\~nez, Ray, and
van Kolck \cite{ORK94,Kol99} who 
constructed a NN potential in coordinate space
based upon $\chi$PT at
next-to-next-to-leading order (NNLO; $\nu=3$).
The results were encouraging and
many researchers became attracted to the new field.
Kaiser, Brockmann, and Weise~\cite{KBW97} presented the first model-independent
prediction for the NN amplitudes of peripheral
partial waves at NNLO.
Epelbaum {\it et al.}~\cite{EGM98} developed the first momentum-space
NN potential at NNLO, and Entem and Machleidt~\cite{EM03} presented the first
potential at N$^3$LO ($\nu = 4$).

In $\chi$PT, the NN amplitude is uniquely determined
by two classes of contributions: contact terms and pion-exchange
diagrams. There are two contacts of order $Q^0$ 
[${\cal O}(Q^0)$] represented by the four-nucleon graph
with a small-dot vertex shown in the first row of Figure~1.
The corresponding graph in the second row, four nucleon legs
and a solid square, represent the 
seven contact terms of ${\cal O}(Q^2)$. 
Finally, at ${\cal O}(Q^4)$, we have 15 contact contributions
represented by a four-nucleon graph with a solid diamond.

Now, turning to the pion contributions:
At leading order [LO, ${\cal O}(Q^0)$, $\nu=0$], 
there is only the wellknown static one-pion exchange, second
diagram in the first row of Figure~1.
Two-pion exchange (TPE) starts
at next-to-leading order (NLO, $\nu=2$) and all diagrams
of this leading-order two-pion exchange are shown.
Further TPE contributions occur in any higher order.
Of this sub-leading TPE, we show only
two representative diagrams at NNLO and three diagrams at N$^3$LO.
The TPE at N$^3$LO has been calculated first by
Kaiser~\cite{Kai01}. All $2\pi$ exchange diagrams/contributions up to
N$^3$LO are summarized in a pedagogical and systematic
fashion in Ref.~\cite{EM02} where 
the model-independent results for NN scattering in peripheral
partial waves are also shown. 

Finally, there is also three-pion exchange, which
shows up for the first time 
at N$^3$LO (two loops; one representative $3\pi$ diagram
is included in Figure~1). 
In Ref.~\cite{Kai99},
it was demonstrated that the 3$\pi$ contribution at this order
is negligible. 

One important advantage of $\chi$PT is that it makes specific
predictions also for many-body forces. For a given order of $\chi$PT,
two-nucleon forces (2NF), three-nucleon forces (3NF), 
\ldots are generated on the same footing (cf.\ Figure~1). 
At LO, there are no 3NF, and
at next-to-leading order (NLO),
all 3NF terms cancel~\cite{Wei90,Kol94}. 
However, at NNLO and higher orders, well-defined, 
nonvanishing 3NF occur~\cite{Kol94,Epe02b}.
Since 3NF show up for the first time at NNLO, they are weak.
Four-nucleon forces (4NF) occur first at N$^3$LO and, therefore,
they are even weaker.

\section{Chiral NN potentials}

\begin{table}[b]
\caption{
$\chi^2$/datum for the reproduction of the 1999 $np$ 
database below 290 MeV by various $np$ potentials.
($\Lambda=500$ MeV in all chiral potentials.)}
\begin{tabular*}{\textwidth}{@{\extracolsep{\fill}}cccccr}
\hline
\hline
 Bin (MeV) 
 & \# of data 
 & N$^3$LO$^a$
 & NNLO$^b$
 & NLO$^b$ 
 & AV18$^c$
\\
\hline
0--100&1058&1.06&1.71&5.20&0.95\\
100--190&501&1.08&12.9&49.3&1.10\\
190--290&843&1.15&19.2&68.3&1.11\\
\hline
0--290&2402&1.10&10.1&36.2&1.04\\
\hline
\hline
\end{tabular*}
\\
\footnotesize
$^a$Reference~\cite{EM03}.\\ 
$^b$Reference~\cite{Epe02}.\\ 
$^c$Reference~\cite{WSS95}.
\end{table}

The two-nucleon system is non-perturbative as evidenced by the
presence of shallow bound states and large scattering lengths.
Weinberg~\cite{Wei90} showed that the strong enhancement of the
scattering amplitude arises from purely nucleonic intermediate
states. He therefore suggested to use perturbation theory to
calculate the NN potential and to apply this potential
in a scattering equation (Lippmann-Schwinger or Schr\"odinger 
equation) to obtain the NN amplitude. We follow 
this philosophy.

Chiral perturbation theory is a low-momentum expansion.
It is valid only for momenta $Q \ll \Lambda_\chi \approx 1$ GeV.
Therefore, when a potential is constructed, all expressions (contacts and
irreducible pion exchanges) are multiplied with a regulator function,
\begin{equation}
\exp\left[ 
-\left(\frac{p}{\Lambda}\right)^{2n}
-\left(\frac{p'}{\Lambda}\right)^{2n}
\right] \; ,
\end{equation}
where $p$ and $p'$ denote, respectively, the magnitudes
of the initial and final nucleon momenta in the center-of-mass
frame; and $\Lambda \ll \Lambda_\chi$. The exponent $2n$ is to be chosen
such that 
the regulator generates powers which are beyond
the order at which the calculation is conducted.

\begin{table}[b]
\caption{
$\chi^2$/datum for the reproduction of the 1999 $pp$ 
database below 290 MeV by various $pp$ potentials.
($\Lambda=500$ MeV in all chiral potentials.)}
\begin{tabular*}{\textwidth}{@{\extracolsep{\fill}}cccccr}
\hline
\hline
 Bin (MeV) 
 & \# of data
 & N$^3$LO$^a$
 & NNLO$^b$
 & NLO$^b$
 & AV18$^c$
\\
\hline
0--100&795&1.05&6.66&57.8&0.96\\
100--190&411&1.50&28.3&62.0&1.31\\
190--290&851&1.93&66.8&111.6&1.82\\
\hline
0--290&2057&1.50&35.4&80.1&1.38\\
\hline
\hline
\end{tabular*}
\\
\footnotesize
$^a$Reference~\cite{EM03}.\\
$^b$See footnote~\cite{note3}.\\
$^c$Reference~\cite{WSS95}.
\end{table}

NN potentials based upon 
$\chi$PT at NNLO~\cite{EGM98,Epe02} are poor in quantitative terms; they
reproduce the NN data below 290 MeV
lab.\ energy with a $\chi^2$/datum of more than 20 (cf.\ Tables 2 and 3).
As shown first by Entem and Machleidt in 2003~\cite{EM03},
one has to go to order N$^3$LO to obtain a NN potential
of acceptable accuracy. For a more recent construction of an
N$^3$LO NN potential, see Ref.~\cite{EGM05}.

For an accurate fit of the low-energy $pp$ and $np$ data, 
charge-dependence is important.
Charge-dependence up to next-to-leading order 
of the isospin-violation scheme 
(NL\O, in the notation of Ref.~\cite{WME01}) includes:
the pion mass difference in OPE and the Coulomb potential
in $pp$ scattering, which takes care of the L\O\/ contributions. 
At order NL\O\, we have pion mass difference in the NLO part of TPE,
$\pi\gamma$ exchange~\cite{Kol98}, and two charge-dependent
contact interactions of order $Q^0$ which make possible
an accurate fit of the three different $^1S_0$ scattering 
lengths, $a_{pp}$, $a_{nn}$, and $a_{np}$.

In the optimization procedure, we fit first phase shifts,
and then we refine the fit by minimizing the
$\chi^2$ obtained from a direct comparison with the data.
The $\chi^2/$datum for the fit of the $np$ data below
290 MeV is shown in Table~2, and the corresponding one for $pp$
is given in Table~3.
The $\chi^2$ tables show the quantitative improvement
of the NN interaction order by order in a dramatic way.
Even though there is considerable improvement when going from
NLO to NNLO, it is clearly seen that N$^3$LO is needed
to achieve an accuracy comparable
to  the phenomenological high-precision Argonne $V_{18}$
potential~\cite{WSS95}.
Note that proton-proton data
have, in general, smaller errors than $np$ data
which explains why the $pp$ $\chi^2$ are always larger.

The phase shifts for $np$ scattering below 300 MeV
lab.\ energy are displayed in Figure~2.
What the $\chi^2$ tables revealed, can be seen graphically
in this figure. The $^3P_2$ phase shifts are a particularly
good example: NLO (dotted line) is clearly poor. NNLO
(dash-dotted line)
brings improvement and describes the data up to about
100 MeV. The difference between the NLO and NNLO
curves is representative for the uncertainty at NLO
and, similarly, the difference between NNLO and N$^3$LO reflects
the uncertainty at NNLO.
Obviously, at N$^3$LO ($\Lambda=500$ MeV, thick solid line)
we have a good description up to 300 MeV. An idea
of the uncertainty at N$^3$LO can be obtained by varying
the cutoff parameter $\Lambda$. The thick dashed line
is N$^3$LO using $\Lambda=600$ MeV. 
In most cases, the latter two curves are not distinguishable
on the scale of the figures. Noticeable differences occur
only in $^1D_2$, $^3F_2$, and $\epsilon_2$ above 200 MeV.

\begin{figure}
\vspace*{-0.5cm}
\hspace*{-1.1cm}
\epsfig{file=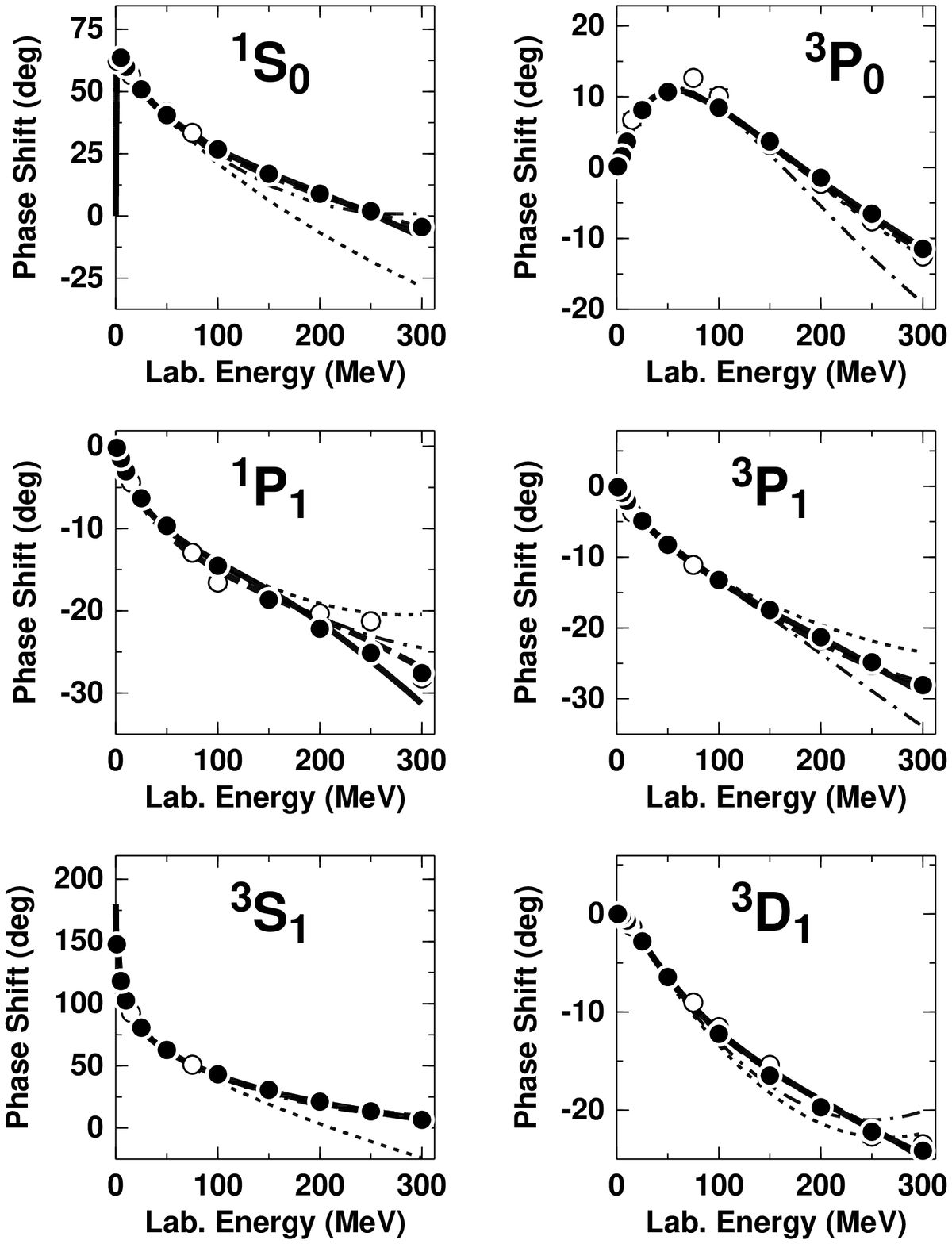,width=73mm}
\hspace*{-1.7cm}
\epsfig{file=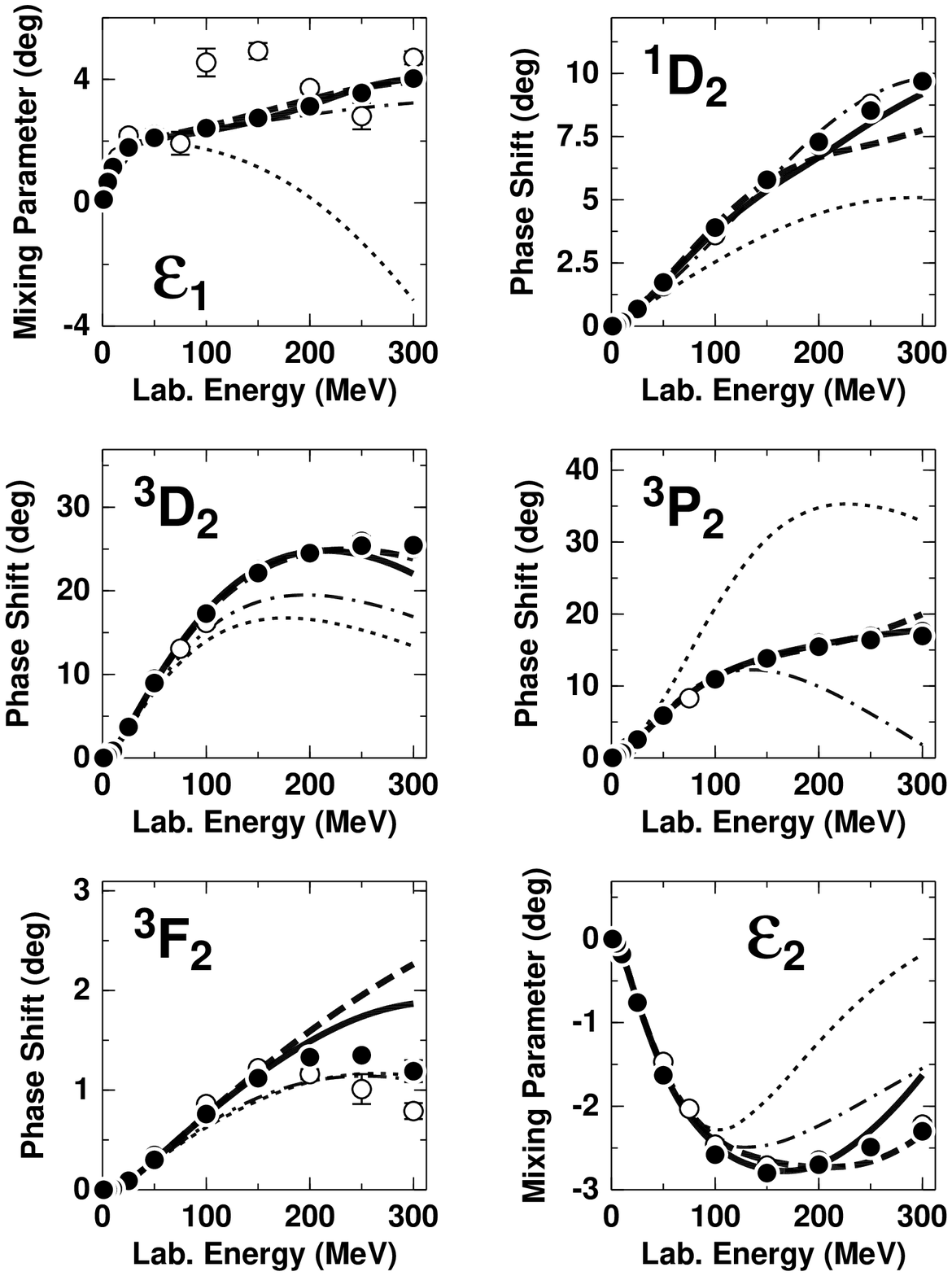,width=73mm}
\vspace*{-1.5cm}
\caption{$np$ phase parameters below 300 MeV lab.\ energy for
partial waves with $J\leq 2$. The thick solid (dashed) line is the result
by Entem and Machleidt~\protect\cite{EM03}
at N$^3$LO using $\Lambda=500$ MeV ($\Lambda=600$ MeV).
The thin dotted and dash-dotted lines are the phase shifts at
NLO and NNLO, respectively, as obtained by 
Epelbaum {\it et al.}~\protect\cite{Epe02} using $\Lambda=500$ MeV.
The solid dots show the Nijmegen multienergy $np$ phase shift
analysis~\protect\cite{Sto93}, and the open circles are the GWU/VPI 
single-energy $np$ analysis SM99~\protect\cite{SM99}.}
\end{figure}

\section{Chiral three-nucleon forces}

As noted before, 
an important advantage of the EFT approach is that it creates
two- and many-body forces on an equal footing.
The first non-vanishing 3NF terms
occur at NNLO and are shown in Figure~1
(row `$Q^3$/NNLO', column `3N Force'). 
There are three diagrams: the TPE, OPE,
and 3N-contact interactions~\cite{Epe02b}.
The TPE 3N-potential is given by
\begin{equation}
V^{\rm 3NF}_{\rm TPE} = 
\left( \frac{g_A}{2f_\pi} \right)^2
\frac12 
\sum_{i \neq j \neq k}
\frac{
( \vec \sigma_i \cdot \vec q_i ) 
( \vec \sigma_j \cdot \vec q_j ) }{
( q^2_i + m^2_\pi )
( q^2_j + m^2_\pi ) } \;
F^{\alpha\beta}_{ijk} \;
\tau^\alpha_i \tau^\beta_j
\end{equation}
with $\vec q_i \equiv \vec{p_i}' - \vec p_i$, 
where 
$\vec p_i$ and $\vec{p_i}'$ are the initial
and final momenta of nucleon~$i$, respectively, and
\begin{equation}
F^{\alpha\beta}_{ijk} = \delta^{\alpha\beta}
\left[ - \frac{4c_1 m^2_\pi}{f^2_\pi}
+ \frac{2c_3}{f^2_\pi} \; \vec q_i \cdot \vec q_j \right]
+ 
\frac{c_4}{f^2_\pi}  
\sum_{\gamma} 
\epsilon^{\alpha\beta\gamma} \;
\tau^\gamma_k \; \vec \sigma_k \cdot [ \vec q_i \times \vec q_j] \; .
\end{equation}  
The vertex involved in this 3NF term is the two-derivative
$\pi\pi NN$ vertex (large solid dot in Figure~1) which we encountered
already  in the TPE contribution to the 2N potential at NNLO.
Thus, there are no new parameters and the contribution
is fixed by the LECs used in NN.
The OPE contribution is
\begin{equation}
V^{\rm 3NF}_{\rm OPE} = 
D \; \frac{g_A}{8f^2_\pi} 
\sum_{i \neq j \neq k}
\frac{\vec \sigma_j \cdot \vec q_j}{
 q^2_j + m^2_\pi }
( \mbox{\boldmath $\tau$}_i \cdot \mbox{\boldmath $\tau$}_j ) 
( \vec \sigma_i \cdot \vec q_j ) 
\end{equation}
and, finally, the 3N contact term reads
\begin{equation}
V^{\rm 3NF}_{\rm ct} = E \; \frac12
\sum_{j \neq k} 
 \mbox{\boldmath $\tau$}_j \cdot \mbox{\boldmath $\tau$}_k  \; .
\end{equation}
The last two 3NF terms involve two new vertices
(that do not occur in the 2N problem), namely,
the $\pi NNNN$ vertex with parameter $D$
and a $6N$ vertex with parameters $E$.
One way to pin down the two new parameters
is to fit them to
the triton and the $^4$He binding energies.
Once $D$ and $E$ are fixed, the results for other
3N, 4N, \ldots observables are predictions.
Results for 3N scattering observables are reported
in Refs.~\cite{Erm05,Wit06}. Spectra of light
nuclei are calculated in Refs.~\cite{Nog04,Nog06}.
Concerning the famous `$A_y$ puzzle', the above 3NF terms
yield some improvement of the predicted nucleon-deuteron
analyzing powers, however,
the problem is not resolved.

One should note that there are additional 3NF terms at NNLO 
due to relativistic corrections ($1/M_N$ corrections) that
have not yet been included in any calculation.
However, there are all reasons to believe
that these contributions will be very small, probably
negligible.
It is more likely that the problem with the chiral 3NF
is analogous to the one with the chiral 2NF:
namely, NNLO is insufficient and for sufficient
accuracy one has to proceed to N$^3$LO.
Two 3NF topologies at N$^3$LO are indicated in Figure~1.
The N$^3$LO 3NF, which does not depend on
any new parameters, is presently under development.

\section{Conclusions}

The EFT approach to nuclear forces is a modern refinement
of Yukawa's meson theory.
It represents a scheme that has an intimate
relationship with QCD and allows to calculate nuclear forces
to any desired accuracy. Moreover, nuclear two- and many-body
forces are generated on the same footing.

At N$^3$LO~\cite{EM03}, the accuracy is achieved that
is necessary and sufficient for microscopic nuclear structure.
First calculations applying the N$^3$LO
NN potential in the no-core shell model 
\cite{NC04,FNO05,Var05},
the coupled cluster formalism
\cite{Kow04,DH04,Wlo05a,Wlo05b,Dea05,Gou06},
and the unitary-model-operator approach~\cite{FOS04}
have produced promising results. 

The 3NF at NNLO is known~\cite{Epe02b} and has had
first successful applications in few-nucleon reactions~\cite{Erm05,Wit06}
as well as the structure of light nuclei~\cite{Nog04,Nog06}. The
3NF at N$^3$LO is under construction.

It may be too early to claim that the nerver-ending story
is coming to an end, but the story is certainly
converging.

\section*{Acknowledgments}

This work was supported in part by the U.S.
National Science Foundation under Grant No.~PHY-0099444.


\begin{thebibliography}{99}
\bibitem{Yuk35} H. Yukawa (1935) 
{\it Proc. Phys. Math. Soc. Japan} {\bf 17} 48. 
\bibitem{Sup56} (1956) {\it Prog. Theor. Phys. (Kyoto) Supplement} {\bf 3}.
\bibitem{TMO52} M. Taketani {\it et al.} (1952)
{\it Prog. Theor. Phys. (Kyoto)} {\bf 7} 45.
\bibitem{BW53} 
K.A. Brueckner {\it et al.} (1953) 
{\it Phys. Rev.} {\bf 90} 699;
{\it ibid.} {\bf 92}, 1023.
\bibitem{Erw61} A.R. Erwin {\it et al.} (1961)
{\it Phys. Rev. Lett.} {\bf 6} 628;
B.C. Magli\'{c} {\it et al.} (1961)
{\it Phys. Rev. Lett.} {\bf 7} 178.
\bibitem{OBEP} (1967) {\it Prog. Theor. Phys. (Kyoto) Supplement} {\bf 39};
R.A. Bryan  and B.L. Scott (1969)
{\it Phys. Rev.} {\bf 177} 1435;
M.M. Nagels {\it et al.} (1978)
{\it Phys. Rev.} D {\bf 17} 768.
\bibitem{Mac89} R. Machleidt (1989)
{\it Adv. Nucl. Phys.} {\bf 19} 189.
\bibitem{Sto94} V.G.J. Stoks {\it et al.} (1994)
{\it Phys.\ Rev.} C {\bf 49} 2950.
\bibitem{Mac01} R. Machleidt (2001)
{\it Phys. Rev.} C {\bf 63} 024001.
\bibitem{Vin79} R. Vinh Mau R 1979
{\it Mesons in Nuclei}, Vol.~I, ed M Rho and
D H Wilkinson (North-Holland, Amsterdam)
p 151.
\bibitem{Lac80} M. Lacombe {\it et al.} (1980)
{\it Phys. Rev.} C {\bf 21} 861.
\bibitem{MHE87} R. Machleidt {\it et al.} (1987)
{\it Phys. Rep.} {\bf 149} 1.
\bibitem{MW88}
F. Myhrer {\it et al.} (1988)
{\it Rev. Mod. Phys.} {\bf 60} 629.
\bibitem{EFV00} D.R. Entem, F. Fernandez, and A. Valcarce
(2000) {\it Phys.\ Rev.\ C} {\bf 62} 034002.
\bibitem{Wu00} G.H. Wu, J.L. Ping, L.J. Teng, F. Wang, and T. Goldman (2000)
{\it Nucl.\ Phys.} {\bf A673} 273.
\bibitem{Wei79} S. Weinberg (1979)
{\it Physica} {\bf 96A} 327.
\bibitem{Wei90} S. Weinberg (1991) {\it Nucl.\ Phys.} {\bf B363} 3.
\bibitem{ORK94}
C. Ord\'o\~nez, L. Ray, and U. van Kolck (1996)
{\it Phys.\ Rev.} C {\bf 53} 2086. 
\bibitem{Kol99} U. van Kolck (1999) {\it Prog.\ Part.\ Nucl.\ Phys.} {\bf 43} 337.
\bibitem{KBW97} N. Kaiser {\it et al.} (1997)
{\it Nucl.\ Phys.} {\bf A625} 758.
\bibitem{EGM98} E. Epelbaum {\it et al.} (2000)
{\it Nucl.\ Phys.} {\bf A671} 295.
\bibitem{EM03} D.R. Entem and R. Machleidt (2003)
{\it Phys. Rev.} C {\bf 68} 041001.
\bibitem{Kai01} 
N. Kaiser (2001) {\it Phys.\ Rev.} C {\bf 64} 057001;
{\it ibid.} {\bf 65} 017001.
\bibitem{EM02} D.R. Entem and R. Machleidt (2002)
{\it Phys. Rev.} C {\bf 66} 014002.
\bibitem{Kai99} 
N. Kaiser (1999) {\it Phys.\ Rev.} C {\bf 61} 014003;
{\it ibid.} {\bf 62} 024001.
\bibitem{Kol94} U. van Kolck (1994) {\it Phys. Rev.} C {\bf 49} 2932.
\bibitem{Epe02b} 
E. Epelbaum {\it et al.} (2002) {\it Phys. Rev.} C {\bf 66} 064001.
\bibitem{Epe02} 
E. Epelbaum {\it et al.} (2002) {\it Eur. Phys.} J. {\bf A15} 543.
\bibitem{WSS95} R.B. Wiringa, V.G.J. Stoks, and R. Schiavilla (1995)
{\it Phys.\ Rev.} C {\bf 51} 38.
\bibitem{note3} Since Ref.~\cite{Epe02} provides only the
$np$ versions of the NLO and NNLO potentials, we have constructed
the $pp$ versions by incorporating charge-dependence
and minimizing the $pp$ $\chi^2$.
\bibitem{EGM05} E. Epelbaum, W. Gl\"ockle, and U.G. Meissner
(2005) {\it Nucl.\ Phys.} {\bf A747} 362.
\bibitem{WME01} M. Walzl {\it et al.} (2001)
{\it Nucl. Phys.} {\bf A693} 663. 
\bibitem{Kol98} U. van Kolck {\it et al.} (1998)
{\it Phys. Rev. Lett.} {\bf 80} 4386. 
\bibitem{Sto93} V.G.J. Stoks, R.A.M. Klomp, M.C.M. Rentmeester, 
and J.J. de Swart (1993) {\it Phys. Rev.} C {\bf 48} 792.
\bibitem{SM99} R.A. Arndt, I.I. Strakowsky, and R.L. Workman
(1999) George Washington University 
Data Analysis Center (formerly VPI SAID facility),
solution of summer 1999 (SM99).
\bibitem{Erm05} K. Ermisch {\it et al.} (2005) {\it Phys.\ Rev.\ C}
{\bf 71} 064004.
\bibitem{Wit06} H. Witala, J. Golak, R. Skibinski, W. Gl\"ockle, A. Nogga,
E. Epelbaum, H. Kamada, A. Kievsky, and M. Viviani (2006)
{\it Phys.\ Rev.\ C} {\bf 73} 044004.
\bibitem{Nog04} A. Nogga {\it et al.} (2004) {\it Nucl. Phys.}
{\bf A737} 236.
\bibitem{Nog06} A. Nogga, P. Navratil, B.R. Barrett, and J.P. Vary (2006)
{\it Phys.\ Rev.\ C} {\bf 73} 064002.
\bibitem{NC04} P. Navr\'atil and E. Caurier (2004) {\it Phys.\ Rev.\ C}
{\bf 69} 014311.
\bibitem{FNO05} C. Forssen, P. Navr\'atil, W.E. Ormand, and E. Caurier (2005)
{\it Phys.\ Rev.\ C} {\bf 71} 044312.
\bibitem{Var05} J.P. Vary {\it et al.} (2005) {\it Eur.\ Phys.\ J.\ A}
{\bf 25} s01 475.
\bibitem{Kow04} K. Kowalski, D.J. Dean, M. Hjorth-Jensen, and T. Papenbrock (2004)
{\it Phys. Rev. Lett.} {\bf 92} 132501.
\bibitem{DH04} D.J. Dean and M. Hjorth-Jensen (2004)
{\it Phys. Rev.} C {\bf 69} 054320.
\bibitem{Wlo05a} M. Wloch {\it et al.} (2005) {\it J.\ Phys.\ G}
{\bf 31} S1291.
\bibitem{Wlo05b} M. Wloch {\it et al.} (2005) {\it Phys.\ Rev.\ Lett.}
{\bf 94} 21250.
\bibitem{Dea05} D.J. Dean {\it et al.} (2005) {\it Nucl.\ Phys.}
{\bf 752} 299.
\bibitem{Gou06} J.R. Gour {\it et al.} (2006) {\it Phys.\ Rev.\ C}
{\bf 74} 024310.
\bibitem{FOS04} S. Fujii, R. Okamato, and K. Suzuki (2004)
{\it Phys. Rev.} C {\bf 69} 034328.
\end{thebibliography}
\end{document}